\documentclass[10pt,twocolumn]{article}          % LaTeX 2e
\usepackage{latexHICSS}                              % LaTeX 2e
\usepackage{times}                               % LaTeX 2e
\usepackage{graphicx}
\usepackage{amsmath}
\usepackage{verbatim}
\usepackage{cite}
\usepackage{fancyhdr}

\begin{document}
%\addtolength{\floatsep}{-5mm}
%\addtolength{\textfloatsep}{-5mm}
%\addtolength{\intextsep}{-2mm}
\def\negspace{\hspace{0mm}}

\title{
\vspace{-.5in}
%\vspace{-37pt}\raisebox{8 pt}{\normalsize\rm 
%preprint; to appear 51th Hawaii International Conference on System Sciences, January  2018, Big Island,  Hawaii. 
%\copyright\ 2017 IEEE. 
%}\linebreak[1] 
 Exploring Cascading Outages and Weather via Processing Historic Data }
\author{
\normalsize{Ian Dobson~~~ Nichelle'Le K. Carrington}\\ [-1mm]
\normalsize{\hspace{-8mm}Kai Zhou~~~~~~~~~ Zhaoyu Wang}\\[1mm]
\normalsize{\hspace{-2mm} ECpE Department,
Iowa State University}\\  [-1mm]
\normalsize{Ames Iowa USA~~ %\\[-1mm]
dobson@iastate.edu}
\and
\normalsize{\negspace Benjamin A. Carreras}\\[1mm]
\normalsize{\negspace BACV Solutions}\\[-1mm]
%\normalsize{Mohawk Drive}\\
\normalsize{\negspace Oak Ridge TN %37831
 USA}\\ [-1mm]
\normalsize{\negspace bacarreras@gmail.com}
%\and
%Nichelle'Le Carrington Kai Zhou Zhouyu Wang\\
%ECpE Department\\
%Iowa State University\\  
%Ames Iowa USA\\[-1mm]
%\normalsize{dobson@iastate.edu}
\and
\normalsize{\negspace Jos\'e M. Reynolds-Barredo}\\[1mm]
\normalsize{\negspace Departamento de F\'{\i}sica}\\[-1mm]
\normalsize{\negspace \hspace{-5mm}Universidad Carlos III %de Madrid}\\[-1mm]
%28911 Leganes, 
Madrid, Spain}\\[-1mm]
\normalsize{\negspace jmrb2002@gmail.com}
}
\maketitle
\fancyhead[c]{\textnormal{\small preprint; to appear at Hawaii International Conference on System Sciences, January  2018, Big Island,  Hawaii}}
\renewcommand{\headrulewidth}{ 0.0pt}
%\fancyfoot[C]{\fontfamily{ptm}\selectfont\fontsize{10}{10}\null}
 \fancyfoot[L]{~\\[-20pt]Preprint \copyright 2017 IEEE. \small Personal use of this material is permitted. Permission from IEEE must be obtained for all other uses, in any current or future media, including reprinting/republishing this material for advertising or promotional purposes, creating new collective works, for resale or redistribution to servers or lists, or reuse of any copyrighted component of this work in other works.}
  \fancyfoot[C]{~ }

\thispagestyle{fancy}

\begin{abstract}
We describe some bulk statistics of historical initial line outages and
the implications for forming contingency lists and understanding which 
initial outages are likely to lead to further cascading.
We use historical outage data  to estimate the effect of weather on cascading 
via cause codes and via NOAA storm data.
Bad weather significantly increases outage rates and interacts with cascading effects, and should be accounted for 
in cascading models and simulations. 
We suggest how  weather effects can be incorporated into the OPA cascading simulation and validated.
There are very good  prospects for improving data processing  and models for the bulk statistics of 
historical outage data so that cascading can be better understood and quantified.
%Further work on discriminating independent outages 
%and propagation of outages via the network during bad weather in both processing methods and models is indicated.
\end{abstract}

\Section{Introduction}

Cascading failure can be defined as a sequence of dependent outages that successively weakens or degrades the power transmission system 
\cite{TaskForcePESGM08}.
Although the power transmission system is carefully designed and operated to be robust to multiple outages, cascading outages that are large enough to cause load shedding and blackouts do occur. The large cascading blackouts that are of the greatest concern are infrequent, but likely enough to have substantial risk \cite{DobsonCH07,HinesEP09,NewmanREL11,CarrerasPS16}.

\looseness=-1
Cascading is the general way that transmission blackouts become widespread and there are many mechanisms that contribute to initial outages or the subsequent propagation of outages.
There are a correspondingly large variety of models, approximations, simulations, and  procedures to assess and mitigate cascading outages \cite{TaskForcePESGM08,TaskForcePESGM11}. One way to evaluate and improve these efforts is validation with observed historical data \cite{CFWGPS16,CarrerasHICSS13,PapicPMAPS16}. There is now much more systematic and automated collection of outage data by utilities, but the challenges of extracting and processing useful information from the data remain.

In this paper, we report on some bulk statistical processing of 14  years of transmission line outage data from a large North American utility to describe initial line outages and to start to explore the effect of weather on cascading.  
Our data-driven analysis of the effect of weather on the bulk statistics of cascading and aspects of our bulk statistical analysis of initial line outages are novel.
Incorporating some of these effects in the OPA (Oakridge-Pserc-Alaska) cascading blackout simulation is also considered (see summary of OPA in section \ref{OPA}).

Instead of working directly with data as in this paper, one can make simulation models
that use or are tuned to typical parameter values.
Several authors have taken this approach to propose models of weather effects in cascading simulations  \cite{Rios02,CiapessoniSG16,CadiniAE17,YaoArXiv17}.

While historical data processing has many advantages, including no modeling assumptions and a very favorable grounding in reality, it should be noted that the grid evolves over 14 years, and that statistical analysis of historical cascades necessarily describes cascading risk averaged over the time period of observation.

%\begin{figure}[h]
%\centering
%\includegraphics[width=3.32in]{WSCCJuly96linetrips.eps}
%\includegraphics{ex5.eps}
%\caption{Cumulative line trips in WSCC July 1996 blackout.
%Time scale is minutes after 14:00 MDT.}
%   \label{WSCCJuly96linetrips}
%\end{figure} 

\Section{Historical outage data and its processing}

The transmission line outage data consists  of \hbox{42\ 561} automatic and planned line outages recorded by a North American utility
over a period of 14 years starting in January 1999 \cite{BPAwebsite}.
The data includes the outage start time (to the nearest minute), names of the buses at both ends of the line, and the dispatcher cause code.
The automatic line outages are identified. 
All this data is standard and routinely collected by utilities.  For example, this data is reported by North American utilities in  NERC's Transmission Availability Data System (TADS) \cite{NERCreport14,BianPESGM14} and is also collected 
in other countries.

%\looseness=-1
Having formed the network model from  the automatic and planned line outages \cite{DobsonPS16}, the analysis of cascading focuses on 
only the automatic outages. There are 10\,942 automatic outages in the data.
The network model has 614 lines, 361 buses and is a connected network.
%One motivation for analyzing only the automatic outages is that cascading focusses on uncontrolled outages; for example, NERC defines cascading as ``the uncontrolled successive loss of system elements triggered by an incident at any location \cite{NERCweb}."

The structure of cascading is that each cascade starts with initial outages in the first generation 
followed by further outages grouped into subsequent generations  until the cascade stops \cite{DobsonEPES17}.
The first step in processing the line outages is to group the line outages  into individual cascades, and then within each cascade 
to group the outages that occur in close succession into generations.
The grouping of the outages into cascades and generations within each cascade is done based on the outage start times 
according to the methods of  \cite{RenCAS08,DobsonPS12}.
We summarize the procedure  here and refer to  \cite{DobsonPS12} for the details.
The grouping is done by looking at the gaps in start time between successive outages.

If successive outages have a gap of one hour or more, then the outage after the gap starts a new cascade.
(That is, suppose $o_1$, $o_2$, ... are the outage start times in their order of occurrence. A gap of more than one hour is defined as a time interval between successive outage start times $o_{i}$, $o_{i+1}$ such that $o_{i+1}-o_{i}\ge 1 $~hour.
The time before the first  outage start time $o_1$ and the time after the last outage start time are also considered to be gaps of more than hour.
Let $g_1$, $g_2$, ... be all the gaps of more than one hour in their order of occurrence.
Then cascade number $k$ is defined to be all the outages that have start times between the gaps $g_k$ and $g_{k+1}$.
An alternative and equivalent definition is that a cascade is a maximal series of outages with successive outage start times for which the time difference between successive outage start times  in the series is less than one hour. A cascade may consist of one outage or many outages.)

Within each cascade, if successive outages have a gap of more than one minute, then the outage after the gap 
starts a new generation of the cascade.
(That is, suppose $o_{k,1}$, $o_{k,2}$, ... are the outage start times in their order of occurrence for all the outages in cascade $k$ . A gap of more than one minute is defined as a time interval between successive outage start times $o_{k,i}$, $o_{k,i+1}$ such that $o_{k,i+1}-o_{k,i}> 1 $~minute.
The time before the first  outage start time $o_{k,1}$ in the cascade $k$ and the time after the last outage start time 
in  cascade $k$ are also considered to be gaps of more than minute.
Let $g_{k,1}$, $g_{k,2}$, ... be all the gaps of more than one minute in cascade $k$ in their order of occurrence.
Then generation number $\ell$ of cascade number $k$ is defined to be all the outages in cascade $k$ that have start times between the gaps $g_{k,\ell}$ and $g_{k,\ell+1}$.)
Since the outage times are only known 
to the nearest minute, the  order of outages within a generation often cannot be determined.

This simple method of defining cascades and generations of outages appears to effective 
and has gap thresholds consistent 
with power system time scales since  operator actions are usually completed within one hour and 
fast transients and protection actions such as  auto-reclosing  are completed within one minute.
\cite{DobsonPS12} examines the robustness of cascade propagation with respect to varying these gap thresholds.

%\begin{figure}[h!]
%  \centering
%\includegraphics[width=\columnwidth]{samplecascadeonnetwork8.pdf} 
%\sffamily
%  \begin{tabular}{ccc}
%  \renewcommand{\familydefault}{\sfdefault}
%\scriptsize &\scriptsize outage start time &\\[-3pt]
%\scriptsize transmission line&\scriptsize hour:minute&\scriptsize generation \\\hline
%\scriptsize JOSQUIN\,$-$\,ISAAC&\scriptsize  15:22 & \scriptsize 1 \\[-1pt]
%\scriptsize GIBBONS\,$-$\,DOWLAND&\scriptsize  15:25 & \scriptsize 2 \\[-1pt]
%\scriptsize ISAAC\,$-$\,OCKEGHEM&\scriptsize  15:27 & \scriptsize 3 \\[-1pt]
%\scriptsize DOWLAND\,$-$\,BYRD&\scriptsize  15:37 & \scriptsize 4 \\[-3pt]
%\scriptsize ANON\,$-$\,BYRD&\scriptsize  15:37 & \scriptsize 4 \\[-1pt]
%\scriptsize OCKEGHEM\,$-$\,DUFAY No 1&\scriptsize  15:49 & \scriptsize 5 \\[-1pt]
%\scriptsize TYE\,$-$\,TALLIS&\scriptsize  15:57 & \scriptsize 6 \\
% \end{tabular}
%   \caption{Example of a cascade of line outages divided into generations and located on the network. The red network lines are the lines that outage. The numbers are the generation number  of the outage  and show the order in which the outages occur. Outages occurring in sufficiently quick succession are in the same generation. The bus names and outage times are changed for confidentiality. Layout is not geographic. }
%  \label{networkcascade}
%\end{figure}

This data processing applied to the 10\,942 automatic outages yields 6687 cascades. Most of the cascades are short: 84\% of the cascades have only the first generation of outages and do not spread beyond these initial outages. It is important for a fair statistical analysis to include the short cascades (even if they are for other purposes not thought of as cascades); the short cascades usually represent a successful case of resilience in which no load is shed. That is, excluding the short cascades would misleadingly bias the results towards the  more damaging cascades that do not stop quickly.
 
The grouping of outages in each cascade into generations allows the initial outages in the first generation to be distinguished from the subsequently cascading outages in the following generations. This is of interest because the mechanisms and mitigations of the initial line outages differ significantly from the interactions between line outages that are involved in the subsequent cascade. 

\begin{figure}[h]
  \centering
\includegraphics[width=\columnwidth]{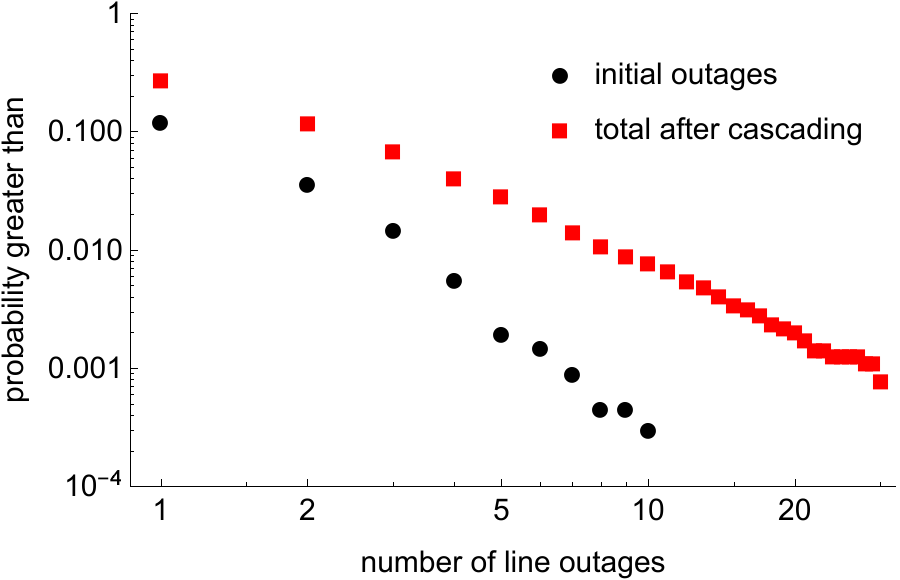} 
   \caption{Probability of more than k outages vs. k for the initial line outages and for the total line outages after cascading}
  \label{alloutagesCCDF}
  \vspace{-5mm}
\end{figure}

Most of the initial line outages are single outages, but there are also multiple initial outages. In other words, there are single, double, triple, etc. contingencies. The probability distribution of the number of initial outages is shown by the black dots in Figure~\ref{alloutagesCCDF}. The distribution of the initial outages is one way of looking at the severity of initial events: 
 12\% of the initial events have more than one outage (the probability of one initial outage is 88\%), 1.5\% of the initial events have more than 3 outages, and 0.2\% have  more than 5 outages.
 
 Cascading increases the probabilities of multiple line outages. The distribution of the total number of outages
 after cascading is shown by the red squares in Figure~\ref{alloutagesCCDF}.
  26\% of the total number of outages have more than one outage, 6.6\% of the total number of outages have more than 3 outages, 2.7\% have more than 5  outages, and 
 0.7\% have 10 or more outages.
 The effect of cascading is progressively larger for the cascades with more outages. For example, while cascading approximately doubles the probability of more than one outage, cascading increases the probability of more than 5 outages by an order of magnitude.
 
The generations of outages in the cascades are analogous to human generations; parents in one generation give rise to children in the next generation. The average propagation per parent $\lambda$ is the total number of child outages in all the generations divided by the total number of parent outages in all the generations. $\lambda$ calculated from our data is 0.28. That is, each outage in a generation will, on average, be followed by 0.28 outages in the next generation.  $\lambda$ quantifies in an overall way how much cascading increases the number of line outages starting from the initial line outages. 

\section{Statistics of initial outages}

We examine the basic statistics of the automatic initial line outages (those outages in the first generation of cascading).
The annual outage frequencies $\mu_1, \mu_2, \mu_3, \ldots, \mu_{614}$ for the 614 lines range from zero outages  to 23 outages per year, with a mean annual frequency $\overline\mu= 0.92$ outages per year. 

The large variation in initial line outage frequency $\mu_i$ in this data has several implications. It is clear that cascading simulations that aim to quantify cascading risk should sample from realistic initial line outage frequencies. One way to accomplish this is to simulate a real power system and use the observed historical frequency of line outages. 
Another way to accomplish this is to understand and model the factors or characteristics that largely determine the frequency of initial line outages so that they can be represented in artificial power system models.
For example, it might be expected that outage frequency has some dependence on line length and other characteristics.
(Our data suggests a mild correlation of 0.3 between outage frequency and line length for lines between 1 and 50 miles long.)
%However, as shown in Figure \ref{outagerateandlength}, there is no apparent correlation between line outage frequency and line length.
Another implication is that it may be difficult to classify the probability of higher order initial outages (when assumed roughly independent) by the number of outages using the order of magnitude of the probabilities \cite{ChenPS05} because the single outages vary so much in their probability.

%\begin{figure}[h]
%\centering
%\includegraphics[width=\columnwidth]{outagerateandlength.eps}
%\caption{Line outage rate versus line length.}
%\label{outagerateandlength}
%\end{figure} 

The data in Figure \ref{alloutagesCCDF} shows a substantial probability of multiple initial outages. The empirical probability of two initial outages is 0.084 with standard deviation 0.0034. To determine whether this can arise from independent single line outages, we suppose that each of the 614 lines has initial outages according to a Poisson process of rate $368/614=0.60$ outages per year and that the Poisson processes for different lines are independent. Then the outages of any of the lines is a Poisson process of rate 368 outages per year, which matches the rate extracted from data in the next paragraph.
Multiple initial line outages in the data require at least 2 line outages to occur at times that are either in the same or adjacent minutes when the times are quantized to minutes. Given the first outage time, this requires the second outage to occur within a 3 minute interval. (For example, if the quantization works by quantizing the time $t$ in minutes to $\lfloor t \rfloor$, the greatest integer number of minutes less than $t$, then $t_1$ and $t_2$ are in the same or adjacent quantized minutes if and only if $\lfloor t_1 \rfloor-1<t_2\le \lfloor t_1 \rfloor+2$.) The probability that the second outage occurs within a given 3 minute interval is $1-\exp[-(3\times 368)/525600]=0.002$. Therefore the probability of multiple initial outages and in particular the probability of 2 initial outages are both bounded above by 0.002. 
Since the empirical probability of 2 initial outages is 0.084, an order of magnitude greater than 0.002, the double initial outages are dependent and cannot be regarded as mainly arising from independent single line outages. It also follows that multiple initial line outages are dependent. A similar claim of dependence for all outages (not just the initial outages)  is established in previous work \cite{ChenPS05,ChenPS06,DobsonPS12}.

\begin{figure}[t]
\centering
\includegraphics[width=\columnwidth]{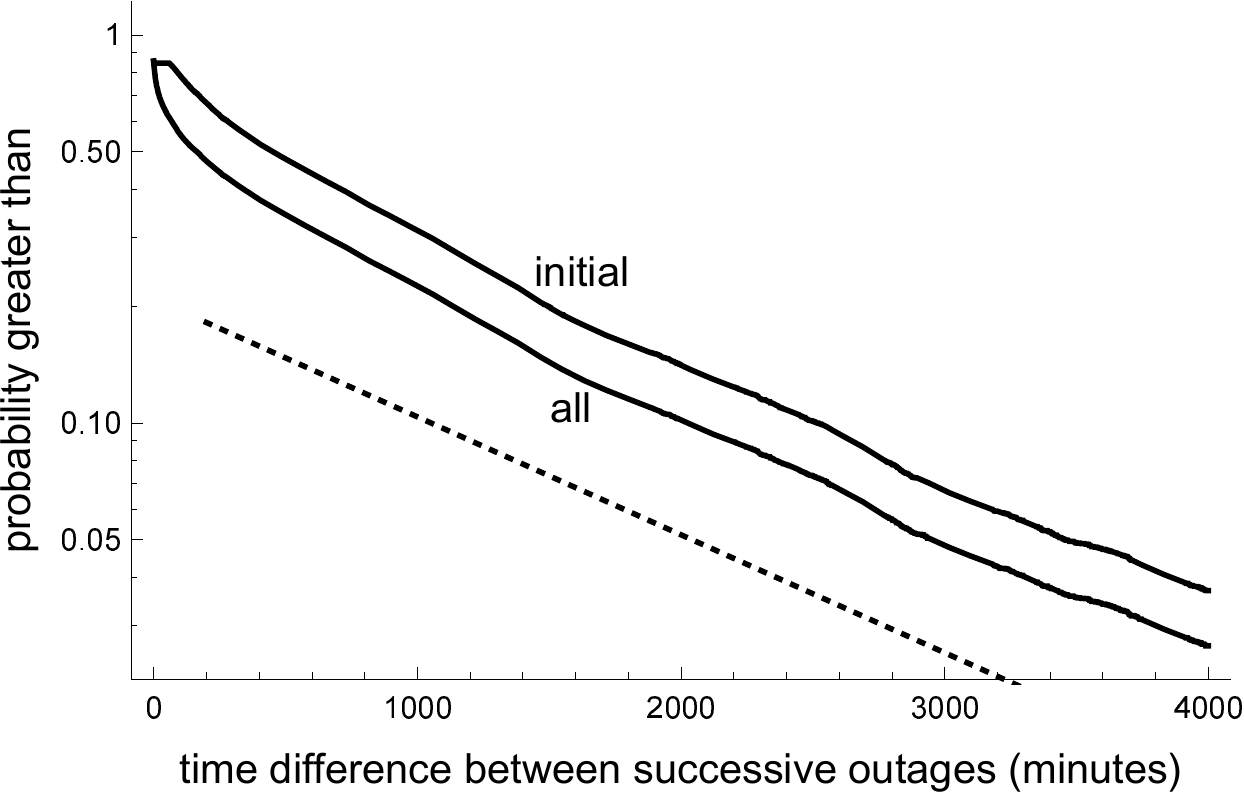}
\caption{Probability distribution (survival function) of time differences between successive initial line outages (upper curve) and all line outages (lower curve).
Note the logarithmic vertical scale.
Dotted straight line has slope corresponding to an exponential distribution of time differences with rate 368 outages per year.}
\label{timedifference}
\end{figure}

The previous paragraph assumes that initial outages are a Poisson process of rate 368 outages per year. 
The assumption that initial outages are a Poisson process is supported by examining the distribution of the logarithm of time differences between successive outage times in Figure \ref{timedifference}. For an exact Poisson process, the time differences follow an exponential distribution with the same rate as the Poisson process, so that the logarithm of the survival function of the time differences is a linear function with the slope of the line determining the rate.
The time differences of both the initial outages and all the outages are approximately linear except for the smaller time differences of order one hour or less.
(The increased, superlinear number of smaller time differences in all the outages may be attributed to cascading. There are no time differences between initial outages more than one minute and less than one hour because of the data processing that defines the start of new cascades. The initial and all outages also show a fraction of outages that have a time difference less than one minute or zero.)
The dotted line in Figure \ref{timedifference} approximates the slope of the time differences except for the smaller time differences of order one hour or less and has a slope corresponding to  368 outages per year. 
The Poisson process model for initial outages is similar to the Poisson process model of blackout start times analyzed in \cite{CarrerasPS16}, except that here we do not consider any adjustment to the Poisson process to account for a slowly varying rate.

\begin{table}
  \caption{Double initial line outages by type}
%  \vspace{2pt}
  \centering
  \vspace{-5pt}
 \begin{tabular}{ccccc}
outaged lines&all&weather&\hspace{-0.5em}no weather&\hspace{-0.8em}random\\
\hline
adjacent&55\%&53\%&55\%&2\%\\
parallel&20\%&11\%&24\%&0\%\\
separated&16\%&26\%&11\%&97\%\\
repeated&9\%&10\%&9\%&1\%
 \end{tabular}
   \label{doubleoutages}
\end{table}

The multiple initial outages have significant spatial dependence. Consider the classification of  initial double outages in Table \ref{doubleoutages}. 
The double outages are either two adjacent lines (lines with exactly one bus in common),
two parallel lines (two buses in common), two separated lines (no common buses), or are repeat outages of the same line.
More than half (55\%) of the double outages are adjacent lines and only 16\% of the double outages are two lines that are separated in the network.
In contrast, randomly sampling double line outages by choosing each of the double lines randomly proportional to their outage frequency yields only 2\%  that are two adjacent lines  and 97\%  that are two lines that are separated in the network.
Combining the adjacent, parallel, and repeated outages shows that 84\% of the initial double line outages form connected subgraphs.

Outages can be caused by line, bus or breaker faults. Line faults are isolated by the breakers at each end of the line so that they usually cause single line outages, while the bus or breaker faults can cause multiple outages because of the substation protection system design.
 Although we do not know any specifics of the substation designs, one likely cause of the high proportion of adjacent double line initial outages is substation breaker schemes that disconnect two lines for certain faults. 

2\% of the initial outages are triple outages and
75\% of these initial triple outages are connected subgraphs.
4\% of the initial outages have 3 or more outages and
68\% of these initial multiple outages are connected subgraphs.
Spatially close components that are assumed to always outage together for cascading failure analysis are called protection control groups \cite{HardimanPMAPS04} or functional groups \cite{ChenPS05}. While there clearly would be some overlap, we do not yet know how exactly the connected subgraphs that form the majority of these initial outages are related to the functional groups that can be applied to approximate the protection system actions. The historical data samples from functional groups, but also samples from rarer or more unusual conditions.

It can also be helpful for risk analysis to find out whether different types of multiple initial outages can cause subsequent outages. Our data shows that separated initial line outages are more likely to trigger subsequent cascading outages: 25\% of separated initial line outages have subsequent outages, while only 16\% of connected initial line outages have subsequent outages. The Mann-Whitney test shows this difference is significant at the 0.01 level (p-value is 0.00355). For momentary initial outages (duration less than one minute) versus  non-momentary initial outages, our data shows little difference in the proportion of subsequent cascading outages: 18\% of momentary outages have subsequent outages, and 15\% of non-momentary outages have subsequent outages. The Mann-Whitney test shows that this difference is not significant at the 0.01 level (p-value is 0.0197).  This suggests that momentary and non-momentary initial outages be treated equally in assessing the risk of further cascading.

The distinction  in the processed line outage data between initial outages and subsequently cascading outages allows us to find out and compare which lines are most involved in these two different processes. The top 10 lines involved in initial outages overlap, but do not coincide with the top 10 lines involved in subsequent cascading; there are 6 lines in common but there are 4 lines in each list that differ. Similarly, the top 20 lines involved in initial outages have 10 lines in common with the top 20 lines involved in subsequent cascading and 10 lines in each list that differ. Similar results were obtained by processing line outage data from a neighboring utility \cite{PapicPMAPS16}. It should be noted that statistically prominent outage problems in an initial portion of the historical data may have been already mitigated.

\section{Effect of weather and other influences via cause codes}
\label{weathercausecode}

The dispatcher outage cause codes allow classification of the cascades of outages into two groups: weather related and non-weather related. (For definiteness when the field and dispatcher causes differ, we do not consider the field cause codes in this analysis.) A cascade of outages is defined  as weather related when at least one of the outages in the cascade has one of  the cause codes ``Weather", ``Lightning", ``Galloping Conductors", ``Ice", ``Wind",  or ``Tree blown".

\def\mystrut(#1,#2){\vrule height #1pt depth #2pt width 0pt}   
\begin{table*}[]
\centering \caption{Some general dependencies of initial outages and average propagation}
\label{tabledependencies}
\centering
\begin{tabular}{cccccl}
%&\multicolumn{2}{|c|}{$d_{\rm mean}^{\rm bus}$}&$d_{\rm maxspread}^{\rm bus}$\\
%&\multicolumn{2}{|c|}{ Between successive generations}&Max from initial\\
equivalent annual&propagation&\multicolumn{3}{c}{$N= $ number of outages in cascade}\\
cascade rate&$\lambda$&$P[N>1]$&$P[N>5]$&$P[N>10]$&\ \ \ \ \ CAUSE\\\hline
\mystrut(12,0) 478&0.28&0.26&0.027&0.007&{\small ALL OUTAGES}\\[2pt]
101&0.55&0.51&0.096&0.028&{\small WEATHER}\\
377&0.13&0.19&0.009&0.002&{\small NOT WEATHER}\\[2pt]
588&0.31&0.29&0.04&0.009&{\small SUMMER MONTHS}\\
423&0.25&0.24&0.02&0.006&{\small NOT SUMMER MONTHS}\\[2pt]
486&0.36&0.34&0.05&0.010&{\small PEAK HOURS}\\
475&0.25&0.24&0.02&0.007&{\small NOT PEAK HOURS}\\[2pt]
\hline
%\omit&\multicolumn{3}{c}{$\pm$ errors are 95\% confidence intervals}\\
 \end{tabular}\end{table*}

\begin{figure}[h]
\centering
\includegraphics[width=\columnwidth]{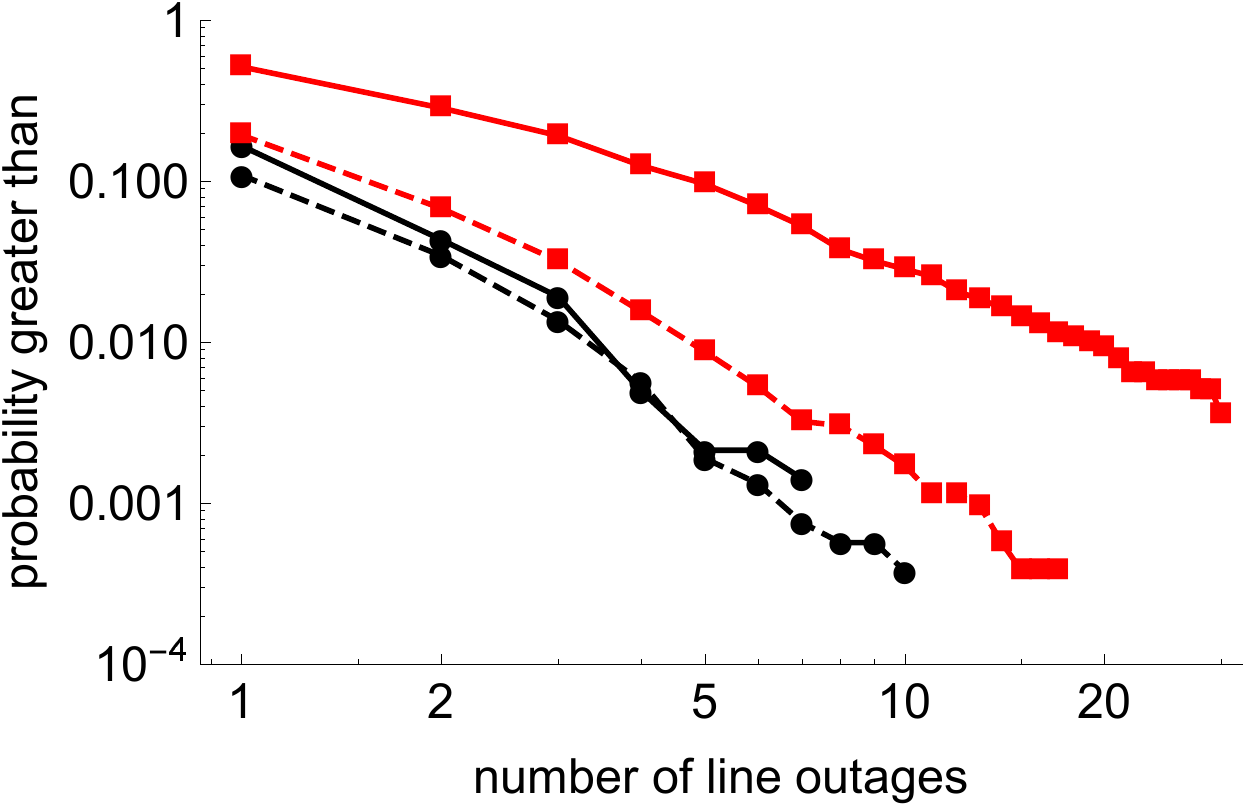}
\caption{Probability distributions of initial (black circles) and cascaded (red squares) outages with weather (solid line) and no weather (dashed line). Weather is determined by cause code.}
\label{weatheroutagesCCDF}
\vspace{-5mm}
\end{figure} 

How the annual cascade rate, average propagation of cascading outages, and cascade size distribution depend on weather are shown in  Table \ref{tabledependencies} and Figure~\ref{weatheroutagesCCDF}.
According to Table \ref{tabledependencies},  only 21\%  (101/478) of the cascades are weather related. 
This implies that only 21\% of the initial outages occurred in a weather-related cascade.
And Figure~\ref{weatheroutagesCCDF} shows that the distribution of initial outages is similar for weather and non-weather related cascades.
However,  in Table~\ref{tabledependencies}  weather related outages have greatly increased propagation from 0.13 (non-weather related) to 0.55  (weather related)
and in Figure~\ref{weatheroutagesCCDF} there  is a corresponding large difference in the distribution of the total outages in a cascade after the cascading.
That is, a minority of cascades are weather related, but they propagate far more to form larger cascades.

With the method of processing outages into cascades that we use \cite{DobsonPS12}, propagation can arise both from outages causing further outages through interactions in the network (encompassing electrical physics, control systems, and human actions) and through independent outages occurring during the cascade that are similar in mechanism to the initial outages. Note that the processing method defining the subsequent  cascading studiously avoids determining the causes or explicit dependences of further outages and simply accounts for outages that occur within one hour of previous outages \cite{DobsonPS12}.
Indeed \cite{DobsonPS12} states that ``One caution is that it is unknown to what extent exogenous forcing from weather is augmented by additional dependent cascading effects."
Reference \cite{DobsonPS16} analyzes all the outages together, determines the average rate of independent outages,  and proceeds to quantify the contribution of statistically independent outages towards the $\lambda$ measure of average propagation, concluding that 4-6\% of outages are independent and classified as cascading outages.
This seems an acceptable error in the contention that the subsequently cascading outages are dependent outages.
The contention essentially relies on the independent outages occurring at a slow enough rate relative to the typical cascade duration.

\begin{figure}[h]
\centering
\includegraphics[width=\columnwidth]{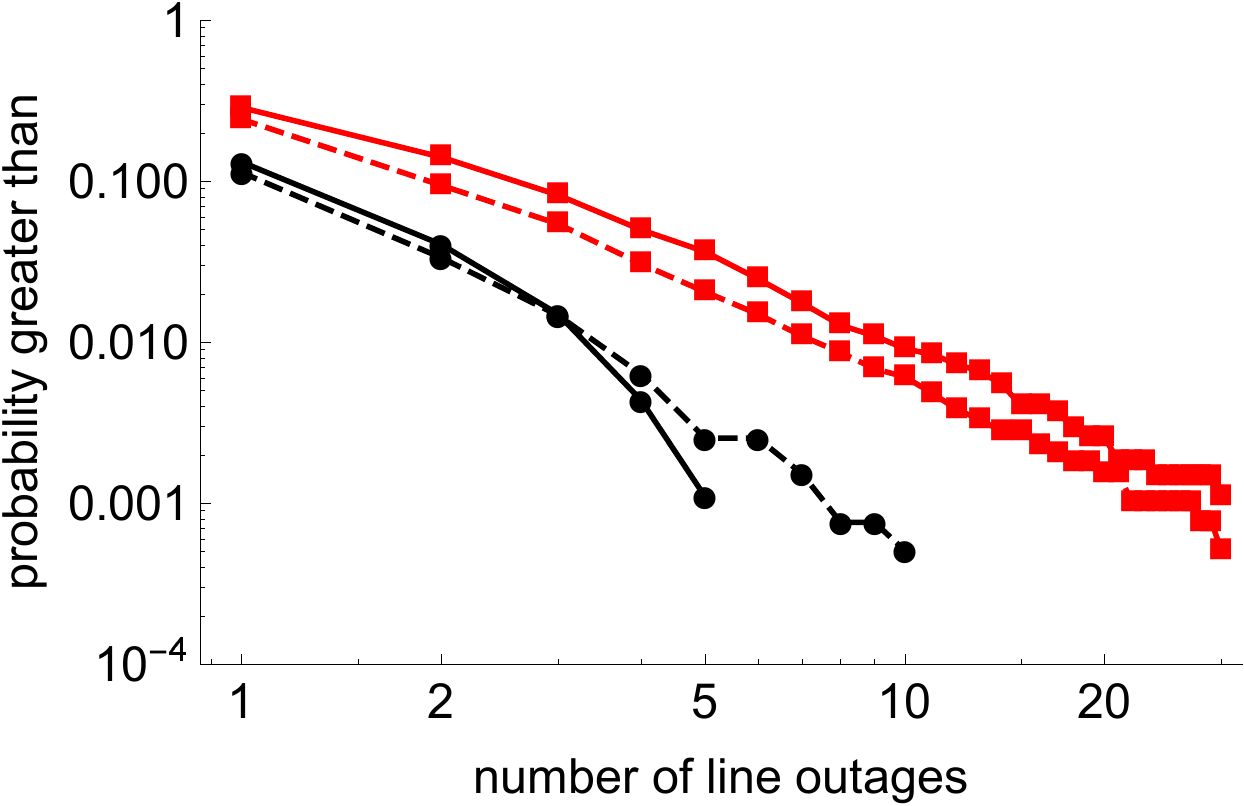}
\caption{Probability distributions of initial (black circles) and cascaded (red squares) outages in summer months (solid line) and remainder of year (dashed line).}
\label{summeroutagesCCDF}
\vspace{-5mm}
\end{figure} 

\begin{figure}[h]
\centering
\includegraphics[width=\columnwidth]{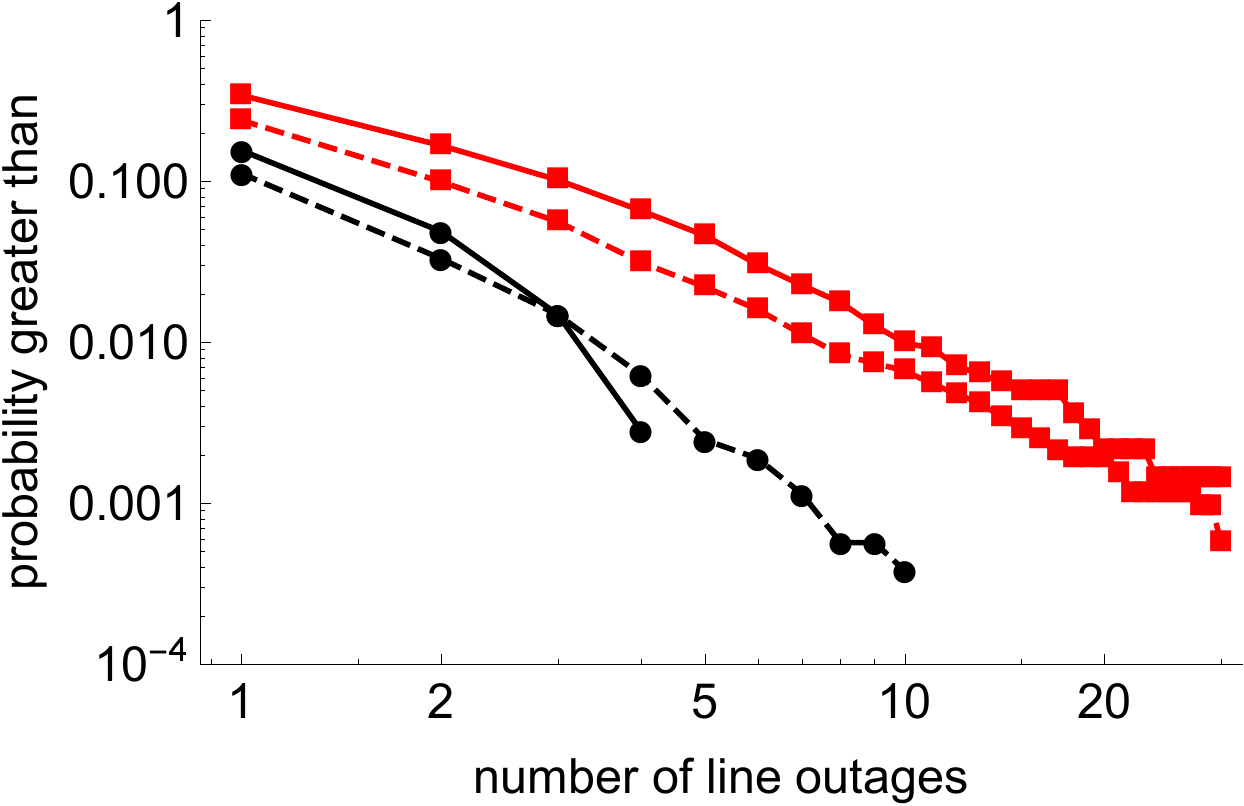}
\caption{Probability distributions of initial (black circles) and cascaded (red squares) outages at peak hours (solid line) and off-peak hours (dashed line).}
\label{peakoutagesCCDF}
\vspace{-5mm}
\end{figure}

However, the same contention for the subset of weather related outages need not have an acceptable error because 
there is a much higher rate of independent outages during bad weather.
The methods of  Section~\ref{weatherNOAA} are not conclusive in this regard, but the results of Section~\ref{weatherNOAA} are 
consistent with this conclusion. More importantly, traditional risk analysis does support  a much higher rate of independent outages during bad weather
\cite{BillintonIEEGTD06,BillintonbookREPS}.
This raises a question of the validity of the method of cascading processing applied to weather related outages when the cascades and propagation are 
interpreted as dependent outages occurring through network interactions.
However, if the concern is simply the number of subsequent outages during a one hour period without regard to cause, then the method could retain some validity for weather related outages. This is the case, for example, when the concern is the total number of outages, regardless of cause,   that the operators have to deal with within a one hour period.

The month and time of day can also be used to classify cascades into the summer peaking months of 
June, July,  August, September, and the remainder of the year, or those cascades that start during the peak hours between 3 pm and 8 pm and cascades starting outside these peak hours.
Table~\ref{tabledependencies} and Figures \ref{summeroutagesCCDF} and \ref{peakoutagesCCDF} show the effects of the summer months and the 
peak hours. (The equivalent annual rate shown in Table~\ref{tabledependencies} is the rate if the condition such as summer months applied all year.) Outages in the summer months of June, July, August, September have  a modestly increased propagation from 0.25 (not summer) to 0.31 (summer). 
Outages in the peak hours between 3 pm and 8 pm have increased propagation from 0.25 (not peak hours) to 0.36 (peak hours). 
Note that the cascades also depend on the initial outages.  Indeed, the data in the summer months shows a 39\% higher cascade rate and a 41\% higher rate of initial outages. 
Overall,  there is a moderate increase in cascade propagation during peak hours and only a small increase in propagation, but an increased rate of initial  outages in the summer months.
However weather effects are larger than either of these factors.

\Section{Effect of weather via NOAA storm data}
\label{weatherNOAA}

\looseness=-1
One problem with analyzing the effect of weather with outage cause codes is that 
  cause codes cannot describe the weather when there is no outage. Therefore the line outage rate during bad weather, a key quantity, cannot be estimated from cause code analysis.
 Also, the outage cause codes are manually entered, rely on subjective judgment about  classifying causes,  and in any case include a sizable proportion (22\% of the dispatcher outage cause codes) of causes ``Unknown".
 One way to address these problems with a different bad weather criterion is to coordinate in time and space the outage data with storm weather records.
 
 The National Oceanic and Atmospheric Administration (NOAA) Storm Events Database is a collection of the occurrence of storm events and other significant weather phenomena recorded by NOAA's National Weather Service from 1950 to present \cite{NOAAdatabase}.
 The NOAA historical storm data records for 1999 to 2013 were obtained for analyzing the storm weather effects influencing our outage data.
The NOAA storm data includes the event type, event start and end time, and the location within the state by county or zone.
The storm event types that we choose to define as a storm  are
``Blizzard", ``Freezing Fog", ``Hail", ``Heavy Rain", ``Heavy Snow", 
``High Wind", ``Ice Storm", ``Lightning", ``Sleet", ``Strong Wind", 
``Thunderstorm Wind", ``Tornado", ``Winter Storm", and ``Winter Weather".

To associate the line outages with the storm data, we map the buses onto the county they are located in, and describe each zone by the main counties it intersects. A line is defined to be in a county if either its sending or receiving end bus is in that county. A line is defined to be in a zone if that zone includes a county that the line is in.  This associates each line with a set of counties. In some cases this set of counties only contains one county. 
A line outage is then classified as a storm outage if it occurs during a storm event in one of the counties in the set of counties associated with that line.
It is straightforward to count the number of storm outages of line number $k$ over the period of observation.
Also, the total storm durations for a county is the sum of the durations of the storms in that county during the period of observation, and the total storm time for line $k$ over the period of observation is computed as the average over the counties of the 
total storm durations for the  counties that the line is in.
Then the line $k$ storm outage rate $R^{\rm storm}_{{\rm line\,}k}=$ (number of storm outages of line $k$)/(total storm time for line $k$).
Finally, the average storm outage rate $R^{\rm storm}={\rm (number~of~lines)}^{-1}\sum_k R^{\rm storm}_{{\rm line\,}k}$.
The non-storm line outage rate and  the  average non-storm line outage rate  $R^{\rm nostorm}$ are computed similarly.

This data processing approximates the average non-storm outage rate as $R^{\rm nostorm}=1.1$ outages per year 
and the average storm outage rate as  $R^{\rm storm}=8.1$ outages per year.
This significant increase in the outage rate during storm weather has important implications for processing historical data and simulating cascading.
First of all, models and simulations of cascading should distinguish and separately consider storm weather and non-storm weather periods.
This conclusion based on cascading historical data is not surprising given the attention to this distinction in conventional power system reliability \cite{BillintonIEEGTD06,BillintonbookREPS} and in \cite{Rios02,CiapessoniSG16,CadiniAE17,YaoArXiv17}.
Secondly, it is also clear from conventional power system reliability that the initial outage rate 
is higher during bad weather \cite{BillintonIEEGTD06,BillintonbookREPS}.
The high outage rate during storms could be primarily  due to increases in the initial outage rate alone or to increases in both the initial outage rate and the propagation.
The distinction matters to mitigation of cascading because the initial outage mechanisms differ from the mechanisms for the propagation of outages through the network.
Thirdly, as already discussed in Section~\ref{weathercausecode}, the significant increase of the average storm line outage rate will also impact the processing of cascading outages into generations. It seems that better high-level models that not only distinguish weather and non-weather conditions but also capture and distinguish the rates of independent and dependent outages are needed. %The use of historical data alongside storm event data will assist in validating the zones of cascading outages along transmission lines that consist of multiple counties and states by tracking the storm's path. This tracking within zones enables the capability of visualizations of the outages as cascading occurs possible.

\begin{figure}[h]
\centering
\includegraphics[width=\columnwidth]{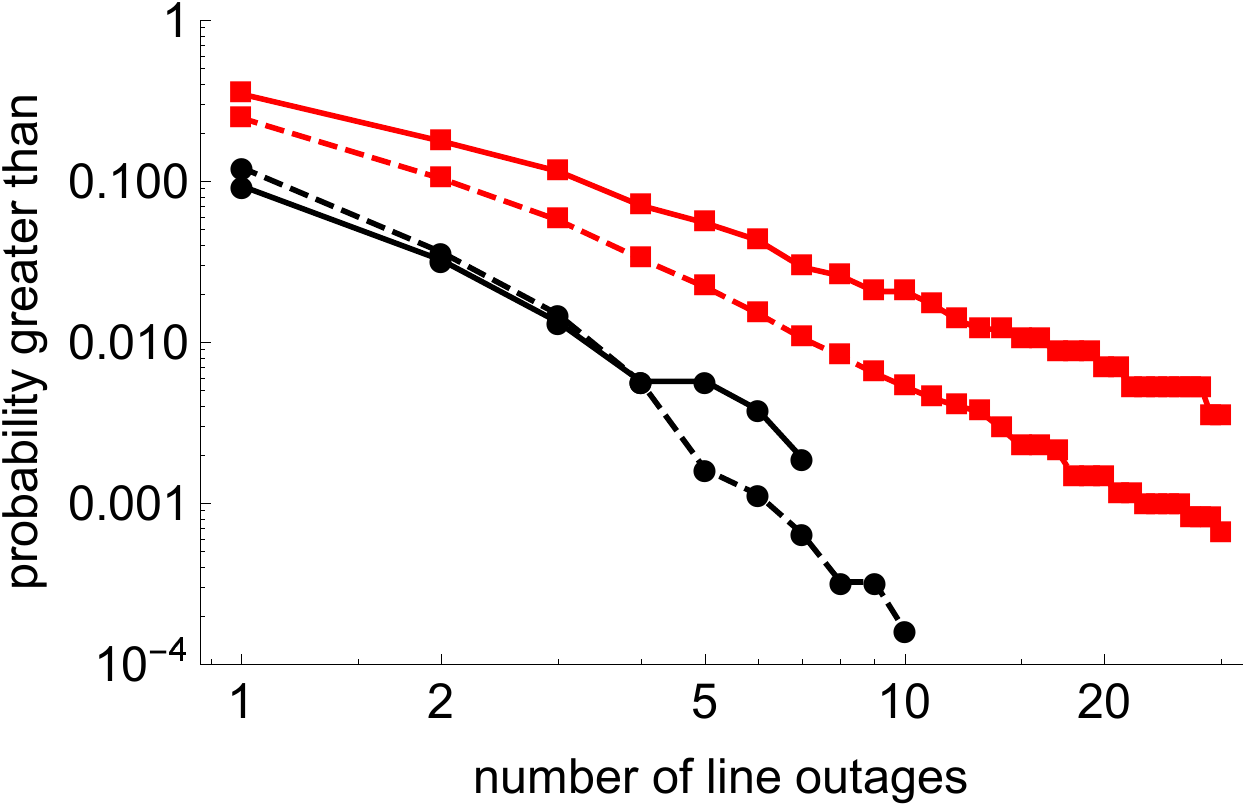}
\caption{Probability distributions of initial (black circles) and cascaded (red squares) outages with storms (solid line) and no storm (dashed line). Storms are  determined by NOAA data.}
\label{stormoutagesCCDF}
\vspace{-5mm}
\end{figure}

The effect of the storm weather on the distributions of initial and total number of cascaded outages  is shown in Figure~\ref{stormoutagesCCDF}.
In comparing Figure~\ref{stormoutagesCCDF} to Figure~\ref{weatheroutagesCCDF}, it should be noted that bad weather or its severity is differently defined by the weather cause codes and the storms in NOAA data.

%\Section{Effect of maintenance (could be included)}

%We study interaction between planned outages and automatic outages in historical data and conclude not much effect.

\Section{Cause and effect in cascading analysis}
\label{cause}

While an attribution of cause for outages is attractive since it often gives possibilities for mitigation,
it should be recognized that the whole notion of detailed cause (and especially single cause) 
for cascading outages can be murky and ill-defined. Causes for initial outages are less problematic, and often 
a single cause or multiple causes can be defined. On the other hand, outages propagating via power system interactions  after 
the initial outages generally depend on the initial system condition, the initial outages, and 
the preceding outages \cite{DobsonEPES17}. 
%Also, outages similar to initial outages can occur independently during the cascade.
%One can address these issues either by starting to build up knowledge of common cause and dependent outage mechanisms
%or by trying to address cascading in general.

To suggest an overall methodological context, we can consider two approaches to cascading: A ``bottom-up" approach 
specifies a particular cause and effect mechanism of cascading, makes a 
detailed model of that mechanism and then tries to get data for that mechanism.
A ``top-down" approach examines available data, at first without regard to detailed cause or mechanism,
and then tries to divide the data into classes of causes or mechanisms.
This paper is top-down and weather is one simple example of a class of mechanisms.
The  bottom-up  
and top-down approaches are complementary.
The bottom-up approach enables understanding and often insight into mitigation of that mechanism, but there are dozens of different mechanisms of cascading, and many of the 
more unusual and complicated events that often occur in blackouts  are hard to approach in this way, and very often there is no data available 
to find the model parameters.
The top-down approach already has the data, and can give a useful overall statistical description and quantification of cascading, 
but gives much less detailed insight and as a consequence work towards mitigation is much less direct.
However, the operators will have to deal with all the outages in real time 
regardless of whether there are detailed cause-effect 
relationships established  or not.
We hope that bottom-up and top-down approaches will gradually converge towards each other as the field progresses
to better realize the full range of possibilities of data-driven reliability analysis.

\Section{Modeling weather effects in OPA}
\label{OPA}

The OPA model \cite{DobsonHICSS01,CarrerasCH04,RenPS08} is a simulation that calculates the long-term risk of cascading blackouts of a power transmission system under the slow, complex dynamics of an increasing power demand and the engineering responses to blackouts. The individual cascades are modeled by probabilistic line overloads and outages in a DC load flow model with linear programming generation redispatch. 
We previously validated OPA on a 1553 bus model of  the Western North American interconnection against observed data with some success \cite{CarrerasHICSS13}. 
Here we start to explore parameterizing the weather effects in OPA on the 1553 bus model by modifying OPA and 
showing the fit with observed data. For details of OPA, parameters,  and the 1553 bus model we refer to  \cite{CarrerasHICSS13}.

\begin{figure}[h]
\centering
\includegraphics[width=\columnwidth]{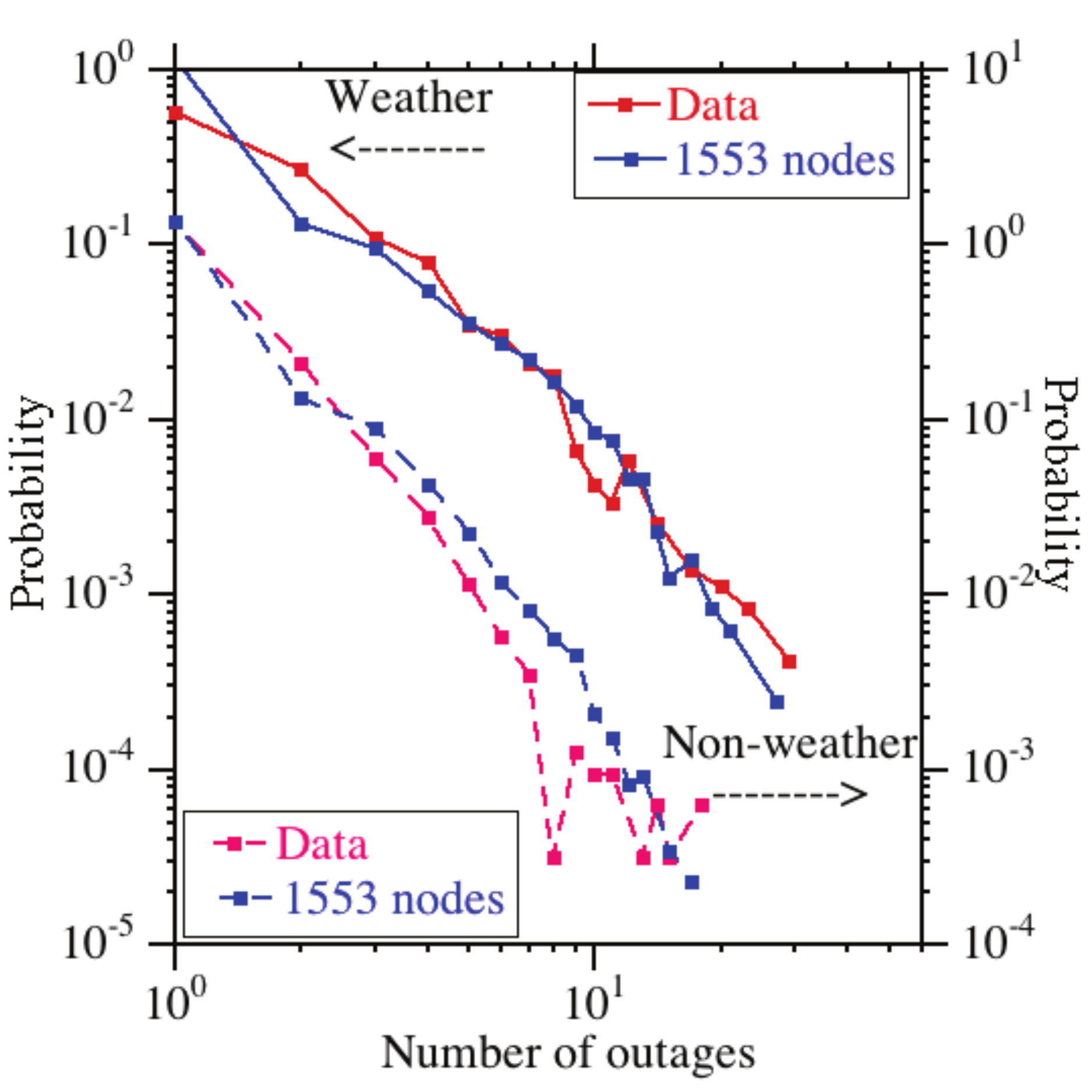}
\caption{PDFs of number of outages in weather and non-weather cascades from OPA simulation and from observed data.}
\label{OPAp5}
\vspace{-5mm}
\end{figure} 

Inspired by the historical data in the previous sections, we  introduce  new OPA parameters $p_w$, $p_5$ and some spatial correlation between multiple initial outages.
$p_w$ is the probability that in a given day the weather is the cause of outages.
$p_5$ is  the probability that an outage produced by weather will happen in a given iteration of the cascade  on the days that weather is a factor. The spatial correlation of the multiple initial outages is introduced by first using OPA parameter $p_0$ to determine some initial line outages. Once initial line outages have been calculated, we go through the adjacent lines and probabilistically determine their failure. Then this process is repeated once. 
The result is that the initial random line outages are sometimes augmented with one or more adjacent lines.

To reexamine the previous fit with the observed data in \cite{CarrerasHICSS13} with the new parameters, we take $p_w=0.25$, which is close to the 21\% proportion of weather cascades 
estimated in section \ref{weathercausecode}. We are not yet able to estimate $p_5$ from data, but $p_5 = 0.0002$
gives a good  match of the cascades with the observed distribution of the number of outages in weather and non-weather cascades as shown in Figure \ref{OPAp5}.
(For $p_5 <0.0002$ the weather driven cascades tend to be too short, and for $p_5 >0.0002$ the slope of the weather PDF from the OPA results tends to decrease relative to the slope of the PDF of the data.)

\begin{figure}[t]
\centering
\includegraphics[width=\columnwidth]{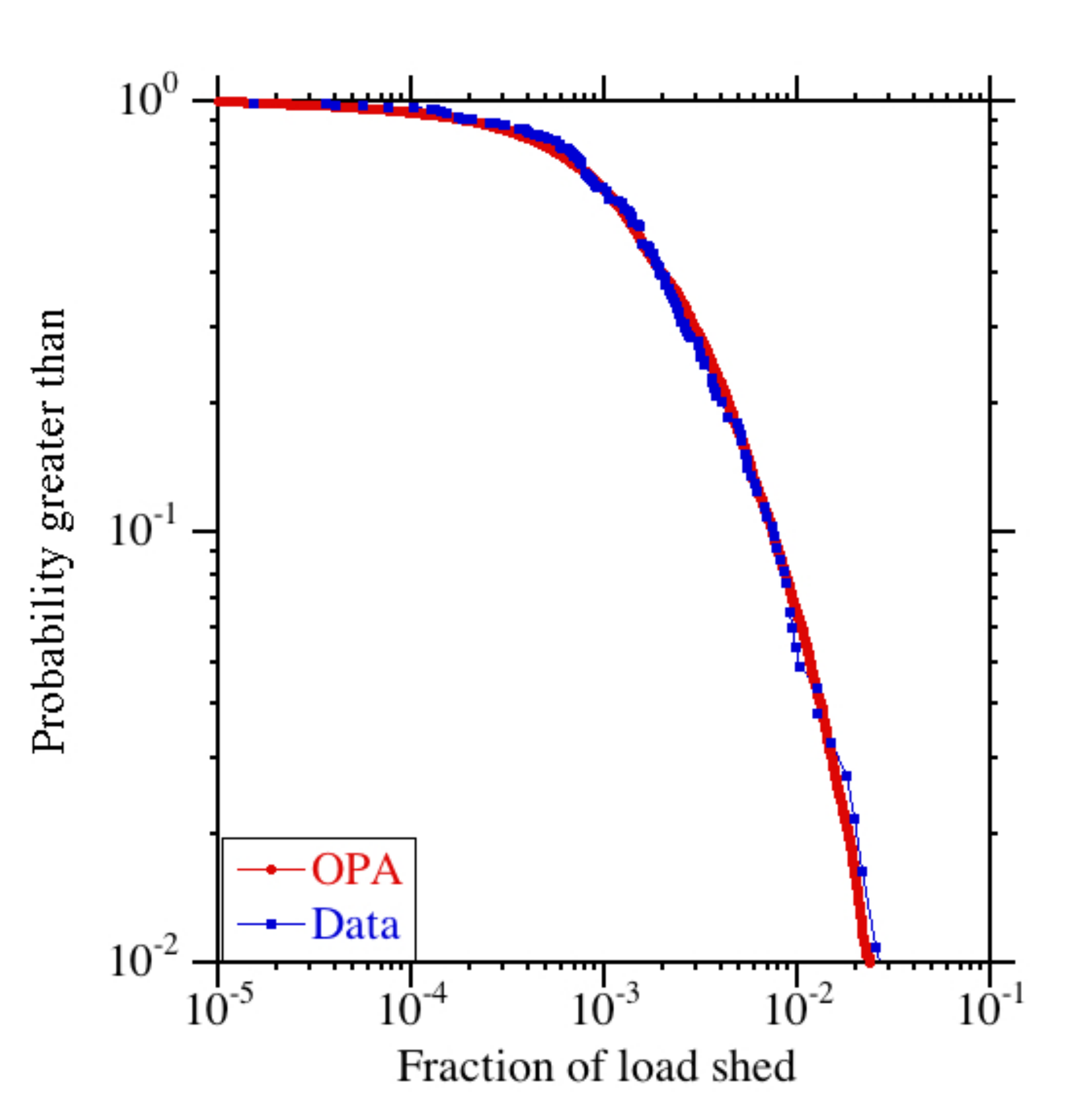}
\caption{Survival function of fractional load shed for OPA and for Western interconnection historical data from the North American Electric Reliability Corporation.}
\label{OPArank}
\vspace{-5mm}
\end{figure}

\begin{figure}[h]
\centering
\includegraphics[width=\columnwidth]{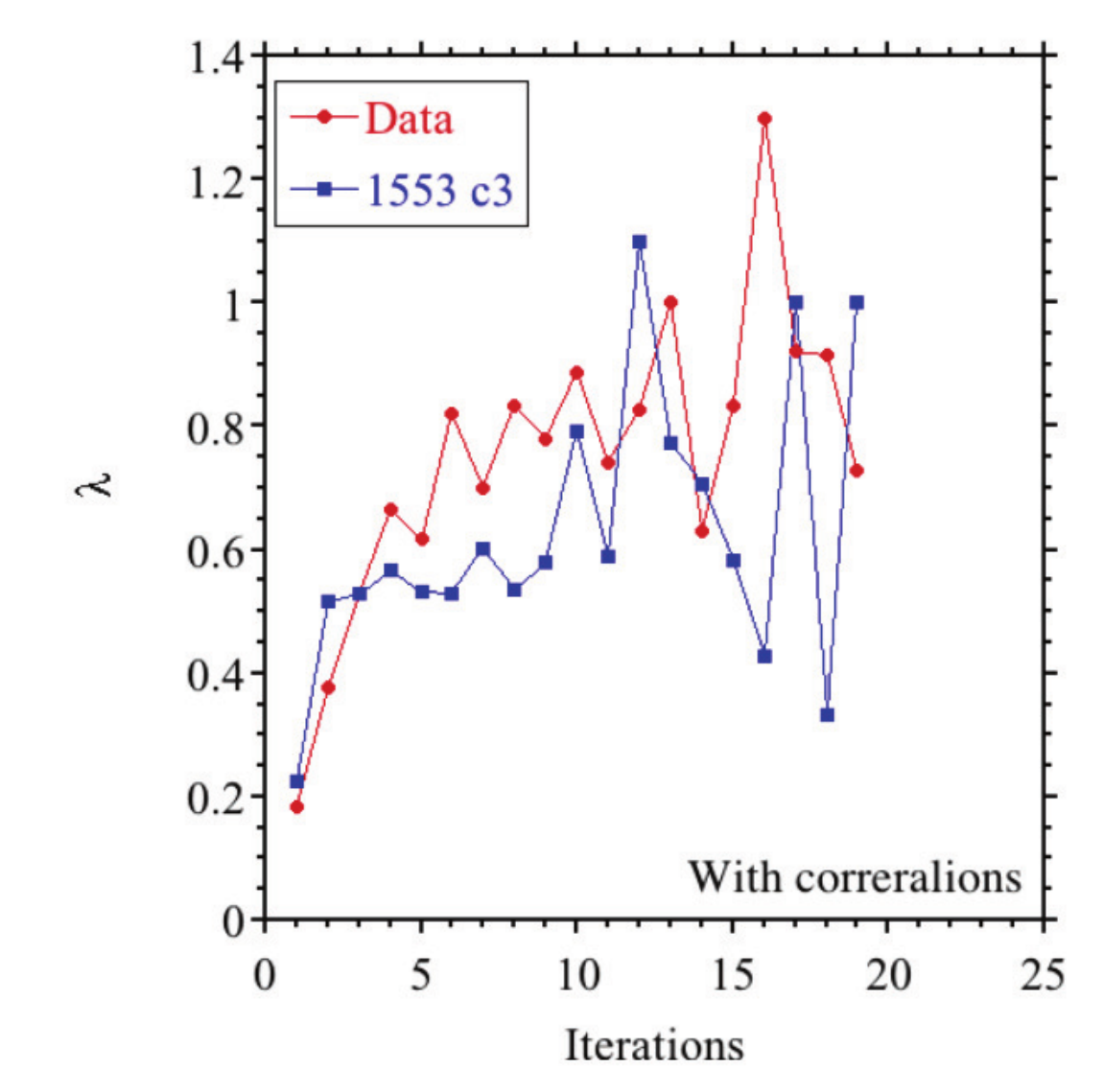}
\caption{Propagation $\lambda$ at each generation (iteration) of cascading from OPA and for the historical data of the paper.}
\label{OPAlambdaall}
\vspace{-5mm}
\end{figure}

With these parameters, in Figure \ref{OPArank}  there is a good match between OPA and the historical data for the distribution of load shed, and in 
Figure~\ref{OPAlambdaall} there is a match for the propagation $\lambda$ in each generation of cascading that improves upon the match in \cite{CarrerasHICSS13}. 
These results suggest that weather effects can be included in OPA and validated against the observed data. Future work should find a way to estimate $p_5$ from the historical data instead of calibrating $p_5$ with the distribution of the number of outages in weather cascades.

\Section{Conclusion}

In this paper we start to explore processing historical outage data to characterize  initial outages and subsequent cascading propagation and determine the effects of weather on cascading.
Although only one 14 year data set from a large North American utility is analyzed and our specific conclusions are of course limited to that data set, 
similar data is routinely collected by many utilities worldwide, so that it is straightforward, given access to the data, to apply the data processing methods broadly.

A simple processing method based on outage timing allows us to distinguish the initial outages from the subsequent cascading. Most of these initial outages are single outages that do not have following outages. However, the data also shows significant numbers of multiple initial outages and initial outages that cascade further. The initial outages have considerable variation in outage frequency, are dependent, and the multiple outages tend to occur in adjacent lines. 
However the separated initial multiple outages have more of a tendency to cascade further. Momentary outages appear to cascade further at a similar rate as longer outages.
As might be expected from the differing mechanisms involved in initial and propagating outages,
the lines most involved in initial outages have some overlap with, but do not coincide with those most involved with subsequent cascading.
The bulk statistics of historical initial outages can inform the contingency lists for risk-based or deterministic cascading studies.

The effects of weather on historical cascading outage data are studied by means of the weather-related dispatcher cause codes in outage data 
and NOAA storm data. 
A minority of cascades are weather-related, but using the processing methods of the paper, show a significantly increased propagation from the initial outages
and a significantly greater outage rate. This suggests that, following traditional power systems risk analysis,  
cascading models and analysis will need to somehow define and acknowledge bad weather and good weather regimes.
An increased outage rate during bad weather is confirmed by traditional power system risk analysis, but its interaction with cascading propagation remains unclear. 
In particular, the increased outage rate does not allow the increased propagation to be mostly attributed to propagation of cascading via network effects 
because of limitations of the processing method. New bulk cascading models and data-processing methods are needed for bad weather conditions.
Peak hours and peak months of operation show less impact on cascading propagation than bad weather, but there is a higher rate of cascades during these peak conditions.

Historical outage data is very valuable for validating and calibrating simulations of cascading outages. 
The OPA model of long-term cascading risk is one of the few simulations with some validation with bulk statistics of historical data \cite{CarrerasHICSS13}.
We have started to represent weather effects in OPA and extend the validation to this case.

Our bulk statistical data processing methods for historical outage data and NOAA data are initial approaches that are subject to future improvements.
However, our results already show the value of this processing for understanding and quantifying key factors in initial outages and subsequent cascading, and the prospects for improved methods and further insights are very good.

%\subsection*{Acknowledgements}

{\bf Acknowledgements:} 
We gratefully thank Bonneville Power Administration and 
the National Oceanic and Atmospheric Administration
%NOAA  
for 
making publicly available the outage and weather data  that made this paper possible.
The analysis and any conclusions are strictly those of the authors and not of
BPA or NOAA.% Bonneville Power Administration or the National Oceanic and Atmospheric Administration.
We gratefully acknowledge support in part from NSF grant 1609080.

%\appendix

%\Section{Probability of $\lambda>\lambda_t$ given $k$ failures}

\end{document}